\documentclass[sigconf]{acmart}

\AtBeginDocument{%
  }

\usepackage{xcolor}
\usepackage{balance}

\usepackage{subcaption}

\copyrightyear{2026}
\acmYear{2026}
\setcopyright{cc}
\setcctype{by}
\acmConference[IDE '26]{3rd International Workshop on Integrated Development Environments }{April 12--18, 2026}{Rio de Janeiro, Brazil}
\acmBooktitle{3rd International Workshop on Integrated Development Environments (IDE '26), April 12--18, 2026, Rio de Janeiro, Brazil}
\acmPrice{}
\acmDOI{10.1145/3786151.3788603}
\acmISBN{979-8-4007-2384-1/2026/04}

\begin{document}

\title{Protecting Private Code in IDE Autocomplete using Differential Privacy}

\author{Evgeny Grigorenko}
\authornote{Authors contributed equally to this research.}
\affiliation{%
  \institution{JetBrains Research}
  \city{Belgrade}
  \country{Serbia}
}
\email{evgeny.grigorenko@jetbrains.com}

\author{David Stanojevi\'{c}}
\authornotemark[1]
\affiliation{%
  \institution{JetBrains Research}
  \city{Belgrade}
  \country{Serbia}
}
\email{david.stanojevic@jetbrains.com}

\author{David Ili\'{c}}
\affiliation{%
  \institution{JetBrains Research}
  \city{Belgrade}
  \country{Serbia}
}
\email{david.ilic@jetbrains.com}

\author{Egor Bogomolov}
\affiliation{%
  \institution{JetBrains Research, TU Delft}
  \city{Amsterdam}
  \country{the Netherlands}
}
\email{egor.bogomolov@jetbrains.com}

\author{Kostadin Cvejoski}
\affiliation{%
  \institution{JetBrains Research}
  \city{Bonn}
  \country{Germany}
}
\email{kostadin.cvejoski@jetbrains.com}

\renewcommand{\shortauthors}{Grigorenko et al.}

\begin{abstract}
Modern Integrated Development Environments (IDEs) increasingly leverage Large Language Models (LLMs) to provide advanced features like code autocomplete. While powerful, training these models on user-written code introduces significant privacy risks, making the models themselves a new type of data vulnerability. Malicious actors can exploit this by launching attacks to reconstruct sensitive training data or infer whether a specific code snippet was used for training. This paper investigates the use of Differential Privacy (DP) as a robust defense mechanism for training an LLM for Kotlin code completion. We fine-tune a \texttt{Mellum} model using DP and conduct a comprehensive evaluation of its privacy and utility. Our results demonstrate that DP provides a strong defense against Membership Inference Attacks (MIAs), reducing the attack's success rate close to a random guess (AUC from 0.901 to 0.606). Furthermore, we show that this privacy guarantee comes at a minimal cost to model performance, with the DP-trained model achieving utility scores comparable to its non-private counterpart, even when trained on 100x less data. Our findings suggest that DP is a practical and effective solution for building private and trustworthy AI-powered IDE features.
\end{abstract}

\ccsdesc[500]{Computing methodologies~Machine learning}
\ccsdesc[500]{Computing methodologies~Natural language processing}
\ccsdesc[500]{Security and privacy}

\keywords{Differential Privacy, LLMs, Code Completion, Fill-in-the-Middle}

\maketitle

\section{Introduction}

Transformer-based large language models (LLMs) \cite{vaswani2017attention} have become an integral part of the modern software development landscape, powering Artificial Intelligence (AI) assistants in many popular IDEs. Tools such as JetBrains AI Assistant~\cite{JetBrainsAI}, Github Copilot~\cite{GitHubCopilot}, Amazon CodeWhisperer~\cite{AmazonCodeWhisperer} use LLMs to provide developers with helpful code suggestions to increase their productivity \cite{ziegler2022productivity}. A significant amount of research has been focused on improving the quality of these code suggestions by exploring different architectures, training pipelines, and curating training data. Within this effort, user data are used both to generate realistic synthetic data and to directly align models with human values, ensuring that they better reflect real-world use cases \cite{MicrosoftResearch2024, kurakin2023harnessing, stiennon2020learning}.

It has been well documented that LLMs can exhibit a property called \textit{memorization} \cite{carlini2021extracting, carlini2022quantifying, lukas2023analyzing, biderman2023emergent}, where, given an appropriate prompt during inference, they can output their training data verbatim. In case of code completion models, sensitive information such as secrets, passwords, or API credentials could potentially be extracted \cite{huang2024your,cheng2025security}.

Recent research shows significant industry-wide privacy and security risks in LLM-based code completion tools. Studies have confirmed the memorization phenomenon, in which models leak the data they were trained on also for code models. For example, \citet{huang2024your} developed a tool that tested prompts from public code repositories and successfully extracted valid credentials. 
Furthermore, \citet{cheng2025security} developed mechanisms that achieved a 99.4\% jailbreak success rate in one leading commercial assistant and a 46.3\%  in another. Furthermore, they demonstrated the extraction of sensitive user data from these tools, including 54 real email addresses and 314 physical addresses associated with user accounts.
Beyond academic research, there are numerous blog posts that highlight these risks, discussing different ways of attacking LLMs or LLMs that have already leaked user personal information \cite{GitGuardian2025,PillarSecurity2025,psystarpsy2025}.

To address these privacy risks, Differential Privacy (DP) \cite{dwork2006differential} has emerged as a technique for training models on sensitive user data, offering strong mathematical guarantees of privacy. However, the practical application of DP in deep learning is notoriously challenging -- one needs to compute per sample gradients which is computationally more expensive as well as adding noise to the gradients to achieve the desired privacy reduces the model utility \cite{bassily2014private, subramani2021enabling}. These technical hurdles have historically limited the scale of DP-trained models, often constraining them to hundreds of millions of parameters rather than billions, typical of their non-private counterparts \cite{anil2021large,li2021large,berrada2023unlocking, ghalebikesabi2023differentially}.

Although recent breakthroughs have produced the first billion-parameter general language model pre-trained with DP \cite{sinha2025vaultgemmadifferentiallyprivategemma}, this progress has yet to extend to the specialized domain of code generation. To our knowledge, no code completion LLM trained with formal DP guarantees has yet been developed, representing a critical gap in building secure, privacy-preserving developer tools.

In this work, we tackle the privacy challenge in LLMs by developing and evaluating, to our knowledge, the first code completion LLM specifically designed for safe integration into IDEs using formal DP guarantees. We apply DP to fine-tune a \texttt{Mellum} \cite{Mellum-4b-base} model, and our empirical results demonstrate a highly favorable privacy-utility trade-off. Although our final privacy budget ($\epsilon \approx 30$) is higher than that often targeted in theoretical work ($\epsilon \le 10$), we show that it provides powerful and practical protection in this real-world IDE scenario. This conclusion, that even large privacy budgets $\epsilon$ hinder the performance of the membership inference attack (MIA) --- methods that given an example $x$ predict if it was used for training or not --- has also been shown in previous work \cite{balle2022reconstructing, kaissis2023bounding, ziller2024reconciling}. Our DP-trained model successfully handles MIA, reducing the attacker's advantage to that of a random guess. This robust practical defense is achieved with only a minimal impact on the model's utility.

The remainder of this paper is structured as follows. We review related work in Section~\ref{sec:related_work}. Section~\ref{sec:background_methodology} provides technical background and details our methodology, threat model, and DP-based defense. We describe the experimental setup in Section~\ref{sec:experimental_setup} and present our empirical results in Section~\ref{sec:results}. Finally, we conclude and discuss future work in Section~\ref{sec:conclusion}.

\section{Related Work}
\label{sec:related_work}

\subsection{Differential Privacy}

Since its introduction, DP has become one of the established mathematical frameworks that provide formal privacy guarantees \cite{dwork2006differential, dwork2014algorithmic}. The development of the differentially private Stochastic Gradient Descent (DP-SGD) method \cite{song2013stochastic,abadi2016deep} made it possible to train neural networks in a differentially private way. Since then, training language models with DP has been an active area of research in both fine-tuning \cite{yu2021differentially, zhang2023dpzero, wu2023privately} and prompting \cite{duan2023flocks, wu2023privacy, amin2024private}.

Most recently \citet{mckenna2025scaling} proposed a new scaling law for differentially private language models, showing that the training dynamic of DP training is significantly different from traditional training and the scaling laws are different. Using these laws \citet{sinha2025vaultgemmadifferentiallyprivategemma} trained the first 1B parameter language model from scratch.

In contrast to the previous work, we focus on supervised fine-tuning of a large pre-trained code completion model on private data. This could potentially provide the possibility of personalization of code suggestions for a particular user or project, while maintaining the security and privacy of the training data.

\subsection{Membership Inference Attacks}

Membership Inference Attack aims to determine whether a specific data record was used to train a target model \cite{shokri2017membership}. This poses a significant privacy threat, as a successful attack can reveal that sensitive information, such as a private code snippet, a medical record, or personal correspondence, was part of the model's training set. Early MIAs often operated in a black-box setting, taking advantage of the principle that models tend to be more confident or exhibit lower loss on the samples in which they were trained \cite{yeom2018privacy, salem2019ml}. More powerful attacks involved training multiple "shadow models" to mimic the target's behavior, thus creating a meta-dataset to train a dedicated attack classifier \cite{shokri2017membership}.

With the rise of LLMs, MIA techniques have grown in sophistication to address the unique properties of these large-scale models. A prominent line of work involves reference-based attacks, which calibrate the membership signal by comparing the target model's output against one or more reference models. The Likelihood Ratio Attack (LiRA), for instance, computes the ratio of likelihoods to distinguish members from non-members with high accuracy \cite{carlini2022quantifying}. These modern attacks often target the phenomenon of \textit{memorization} rather than the simple overfitting \cite{carlini2021extracting}. LLMs can memorize and reproduce parts of their training data, even while demonstrating strong generalization capabilities, making them vulnerable even when not overtly overfit \cite{fu2024miatuner}. Other recent approaches have explored different signals, such as comparing a sample's likelihood to that of its semantic neighbors \cite{mattern2023membership} or analyzing token-level probabilities to detect pre-training data \cite{shi2024detecting}.

The demonstrated effectiveness of these attacks highlights that LLMs trained with private data, such as proprietary source code, represent a significant security vulnerability. This underscores the need for robust defense mechanisms with formal privacy guarantees. Although various defenses such as regularization or confidence masking have been proposed, they often offer unreliable protection \cite{hu2022membership}. Our work addresses this gap by employing DP, a principled framework designed to prevent the model from memorizing individual training examples, thereby providing a strong mathematical defense against the threat of membership inference.

\section{Background and Methodology}
\label{sec:background_methodology}

Our methodology adapts a large publicly trained language model to a private internal dataset while providing formal privacy guarantees using DP.

As starting point we take our in-house \texttt{Mellum} 4B model, a powerful LLM pre-trained on a vast corpus of public source code and equipped with Fill-in-the-Middle (FIM) capabilities (given prefix and suffix as context a LLM is trained to generate the code in the middle) \cite{bavarian2022efficient}. We then fine-tune this model using sensitive data from our internal code repositories, which constitute the private dataset we aim to protect.

For the fine-tuning process, we employ \textit{Low-Rank Adaptation (LoRA)} \cite{hu2022lora}, a parameter-efficient fine-tuning (PEFT) technique. Instead of updating all the weights of the model, LoRA freezes the pre-trained parameters and injects a small number of trainable, low-rank matrices into the model architecture. This approach significantly reduces the number of trainable parameters.

The choice of LoRA is particularly advantageous for training with DP-SGD. By reducing the number of trainable parameters, the dimensionality of the gradients we compute in each step is much smaller. Consequently, less noise must be added to the gradients to achieve a specific privacy budget ($\epsilon$), making privacy-preserving training more efficient and helping to maintain the higher utility of the model.

\subsection{LLMs as Code Completion Models}
\label{subsec:code_completion}

Code completion is a cornerstone feature of modern IDEs, designed to enhance developer productivity by suggesting subsequent tokens or entire blocks of code based on the current context. The FIM paradigm extends the functionality of modern code completion models beyond the simple prediction of the next token \cite{bavarian2022efficient}. Instead of being limited to a left-to-right (causal) generation process, FIM-enabled models can infill missing code by conditioning on both a \textit{prefix} (the preceding code) and a \textit{suffix} (the succeeding code).

This is typically achieved by training the model on a specialized objective. The input is formatted with special tokens to delineate the different parts, for example: \texttt{<PRE>prefix<SUF>suffix<MID>}. The model is then trained to generate the code that should replace the \texttt{<MID>} token, conditioned on both the preceding and the following context. It enables developers to: \textit{complete function bodies} after having written the function signature and the return statement; \textit{fill in missing arguments} in a function call; \textit{refactor code} by automatically generating new logic to connect existing parts.

\subsection{Threat Model}
\label{subsec:threat_model}

\begin{figure}[t!]
    \centering
    \caption{An illustration of a Membership Inference Attack (MIA), where an adversary's goal is to determine if a specific data point (e.g., the one containing the avocado) was part of the original private dataset $D$.}
      \includegraphics[width=0.9\linewidth]{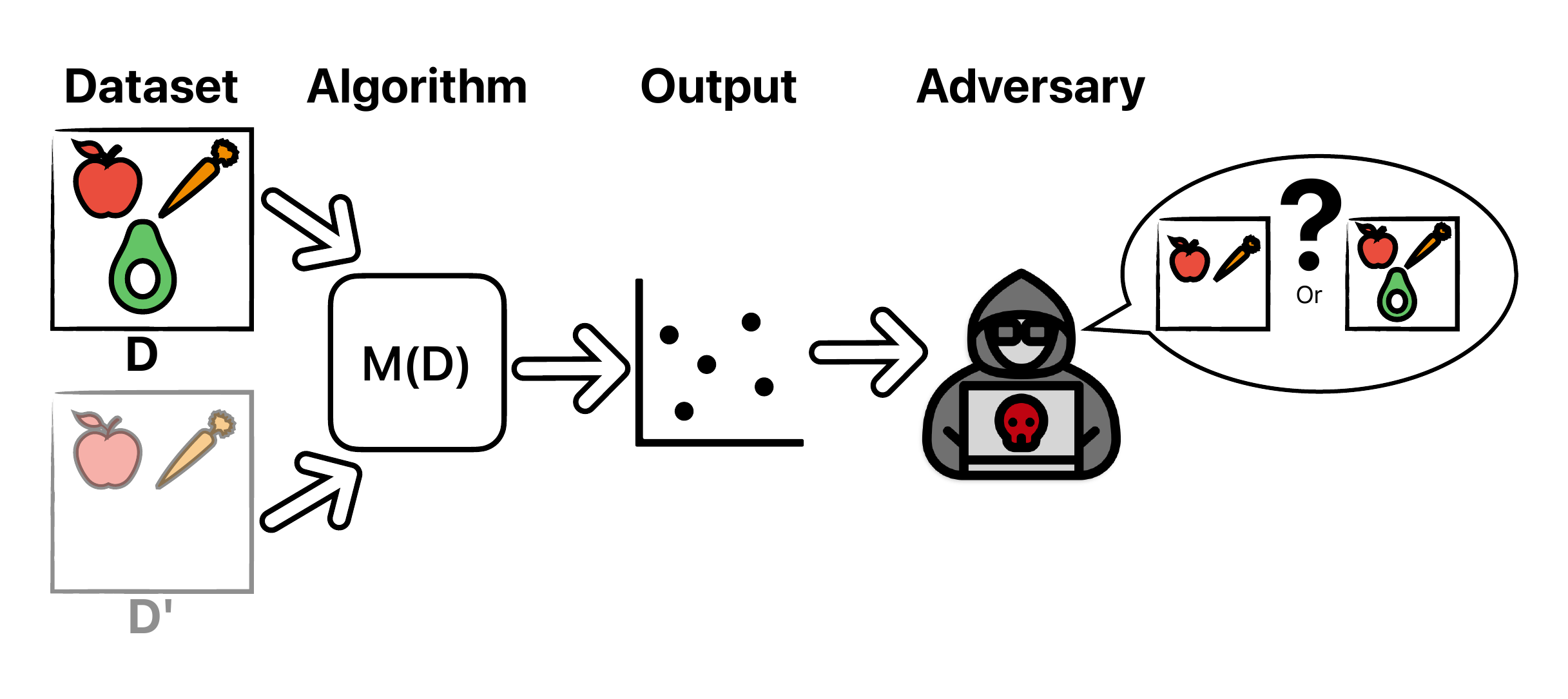}
    \label{fig:mia}
\end{figure}

The integration of LLMs into IDEs introduces a new attack surface where the model itself becomes a potential source of data leakage. Our threat model considers an adversary whose primary goal is to perform a MIA. The adversary's objective is to determine whether a specific, targeted code snippet --- for instance, a proprietary algorithm or a function containing a known vulnerability --- was included in the private dataset used to fine-tune the code completion model.

We assume a realistic "gray-box" access scenario, where the adversary possesses the following capabilities and knowledge:
\begin{itemize}
    \item \textbf{Query Access}: The adversary can interact with the fine-tuned model through an API, providing arbitrary code snippets and obtaining not just the generated completion, but also a confidence score, such as the loss or per-token probabilities.
    \item \textbf{Model Knowledge}: The adversary knows the architecture of the model (e.g., \texttt{Mellum}) and has access to the public, pre-trained base model upon which the private fine-tuning was performed.
    \item \textbf{Target Data}: The adversary has a specific set of code snippets and wants to test their membership in the private training set.
\end{itemize}
Crucially, the adversary does not have access to the model's weights or the complete private fine-tuning dataset. A successful MIA in this scenario would constitute a significant privacy breach, confirming that sensitive internal code was used for training. Figure~\ref{fig:mia} depicts this process. This threat motivates our use of DP, a principled defense mechanism designed to provably limit what an adversary can infer about individual training examples.

\subsection{Defense: Differential Privacy}
\begin{figure}
    \caption{The workflow of Differentially Private Stochastic Gradient Descent (DP-SGD). Per-sample gradients are first computed and then clipped. These clipped gradients are averaged, noise is added , and the final result is used to update the model}
    \centering
    \includegraphics[width=\linewidth]{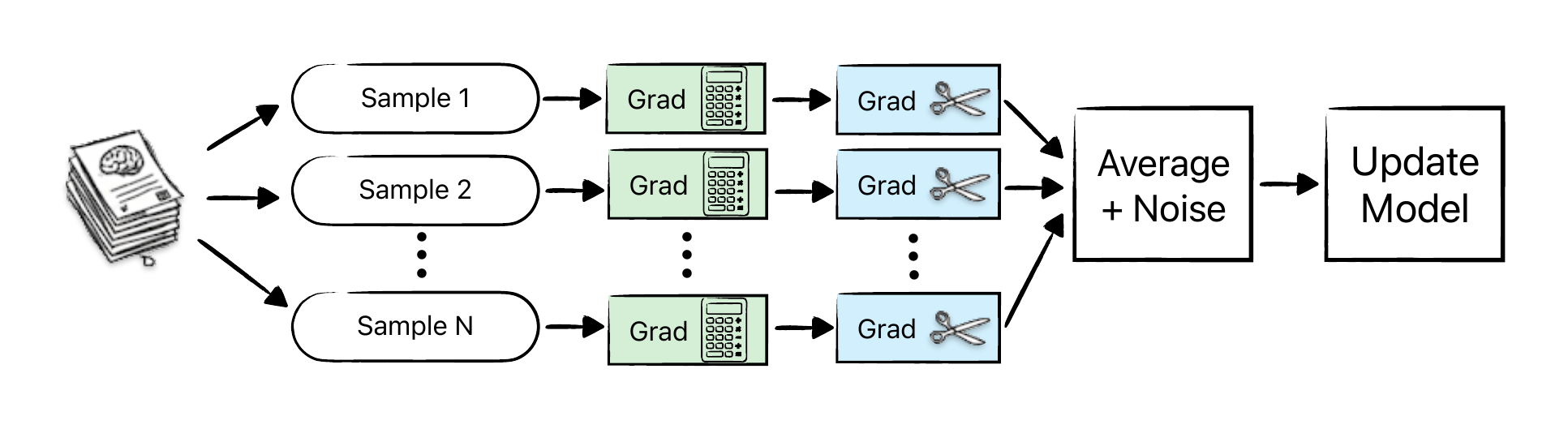}
    \label{fig:dp_sgd}
\end{figure}
Differential Privacy is a mathematical framework that provides a strong provable guarantee of privacy \cite{dwork2006differential}. The core idea is to ensure that the outcome of any analysis is statistically indistinguishable whether or not any single individual's data were included in the dataset. This is achieved by adding carefully calibrated statistical noise to the process.

More formally, a randomized algorithm $\mathcal{M}$ satisfies $(\epsilon, \delta)$ - Differential Privacy if for any two adjacent datasets $D$ and $D'$ (differing by at most one element), and for any set of possible outputs $S$, the following inequality holds,

\begin{equation}
    P(\mathcal{M}(D) \in S) \le e^{\epsilon} \cdot P(\mathcal{M}(D') \in S) + \delta.
\end{equation}

Here, the \textit{privacy budget}, denoted as $\epsilon$, is a small positive real number that measures the \textit{privacy loss}. Smaller $\epsilon$ values correspond to a stricter privacy guarantee, which makes the output of the algorithm on neighboring datasets very similar. The parameter $\delta$, typically a very small number, represents the probability that the strict privacy guarantee might not hold. In order to privately calculate an average salary with DP, for instance, each employee can add random noise from a zero-mean distribution to their actual salary before reporting it. This method protects individual privacy because true salaries are never shared, yet the overall average remains accurate as the noise cancels out in aggregate.

To apply these guarantees to the training of deep learning models, the standard training algorithm, Stochastic Gradient Descent (SGD), is modified into \textit{Differentially Private Stochastic Gradient Descent (DP-SGD)} \cite{song2013stochastic,abadi2016deep}. DP-SGD introduces privacy-preserving steps into the model's weight update process. Instead of simply averaging gradients across a batch of data and updating the model, DP-SGD follows a more careful procedure:

\begin{enumerate}
    \item \textbf{Per-Example Gradient Computation}: First, for a batch of training examples, the gradient of the loss function is computed for each \textit{individual} example.

    \item \textbf{Gradient Clipping}: The influence of each individual data point is limited by \textit{gradient clipping}. The L2 norm of each per-example gradient is calculated, and if it exceeds a predefined clipping threshold, $C$, the gradient is scaled down. This ensures that no single data point can have an outsized impact on the model update.

    \item \textbf{Noise Addition}: The clipped gradients for the batch are then averaged. Before this average gradient is used to update the model, random noise (typically from a Gaussian distribution) is added to it. This step is the core of the privacy guarantee, as the noise masks the precise contributions of the individual data points in the batch.

    \item \textbf{Model Update}: Finally, this noisy average gradient is used to update the model's weights.
\end{enumerate}

Figure~\ref{fig:dp_sgd} depicts the gradient update step of a model with DP-SGD.
Throughout training, a "privacy accountant" tracks the cumulative privacy cost of every update, resulting in a final $(\epsilon, \delta)$ guarantee for the trained model. This process ensures that the model learns general patterns from the data without memorizing specific, private details from individual training examples.

\section{Experimental Setup}
\label{sec:experimental_setup}

\subsubsection*{Models and Datasets}
\label{subsec:models_datasets}

Our experiments begin with the \texttt{Mellum SFT} model \cite{Mellum-4b-base}, a 4-billion-parameter, LLaMA-style Transformer pre-trained on over 4 trillion tokens of public code from multiple programming languages, with a specialization in Kotlin. The model supports a context window of 8192 tokens and is equipped with FIM capabilities, making it highly effective for code completion tasks.
For fine-tuning, we use a private dataset composed of code from our internal Kotlin repositories\footnote{In this work we use only code from our internal JetBrains repository. The model that we developed is used only for research purposes and it is not deployed. Data from commercial users was not used.}. The dataset has $8M$ training instances with a maximum length of 8192.

Model performance is evaluated using an internal Kotlin benchmark dataset that neither model has seen during training.

\subsubsection*{Training Procedure}
\label{subsec:training_procedure}

We train two distinct models to compare their performance and privacy characteristics.

\paragraph{Baseline (Non-DP)} The non-private baseline, which we refer to as \texttt{Mellum Int. Base}, is fine-tuned on the full 8 million instances from our internal dataset.

\paragraph{Differentially Private Model} Our private model, \texttt{Mellum Int. DP}, is fine-tuned on $80K$ instances from the same internal dataset as the \texttt{Mellum Int. Base} is trained on. We use only $80K$ instances because of the early stopping that we have for the privacy budget $\epsilon$. For finding the best hyper-parameter grid search was conducted, LoRA rank $r=\{4, 8, 16, 32, 64\}$ and privacy budget $\epsilon=\{4, 30, 100, 1000\}$. The training is performed using DP-SGD more specifically we use AdamW \cite{loshchilov2017decoupled} as optimizer. The entire training pipeline is based on the Opacus library \cite{opacus-website, yousefpour2021opacus}. The key DP parameters were a noise multiplier of $\sigma = \text{0.2746}$ and a gradient clipping norm of $C = \text{0.5}$. The model was trained for 1 epoch with an effective batch size of $512$, resulting in a final privacy budget of $(\epsilon \approx 30)$. We settled on LoRA rank $r=8$ and privacy budget $\epsilon=30$ since it gave the best trade-off between privacy and utility.

\subsubsection*{Evaluation}
\label{subsec:evaluation}

In order to evaluate the model trained with DP we have to measure the performance on downstream task (utility) as well as the privacy (i.e. resistance to MIA).

\paragraph{Utility Evaluation}
We assess the code generation quality of our models using two metrics: (1) \textsc{\textbf{ChrF++}} --- is an evaluation metric that measures the quality of generated text by calculating an F-score based on the overlap of both character n-grams and word n-grams between the generated output and a reference; (2) \textsc{\textbf{Longest Match (LM) Score}} --- measures the longest consecutive sequence of matching lines between a model’s completion and the ground-truth completion, and scores it using pre-defined thresholds; it was introduced as a multi-line completion counterpart to exact match; (3) \textsc{\textbf{LLM-as-judge}} --- metrics like \textsc{LM} can be to strict therefore we introduce LLM-as-judge metric that evaluates the quality of each completion in interval $[0,1]$.

\paragraph{Privacy Evaluation}
Privacy is evaluated by performing a MIA. The goal of the MIA is to determine if the model can distinguish between data it was trained on (members) versus unseen data (non-members). The attack performance is quantified by the \textit{Area Under the ROC Curve (AUC)}. An AUC of $1.0$ indicates a perfect attack, allowing the adversary to perfectly distinguish members from non-members. An AUC of $0.5$ signifies that the attacker's performance is no better than a random guess, indicating strong privacy.

\section{Results}
\label{sec:results}

\begin{table}[t!]
\small
\caption{DP and non-DP Mellum models, fine-tuned with LoRA ($r=8$) on internal data, were evaluated against the baseline (\texttt{Mellum SFT}) on internal Kotlin benchmark. \texttt{Mellum Int. Base} is trained on $8M$ and \texttt{Mellum Int. DP} on $80K$ instances.}
\centering
\begin{tabular}{lcccccc}
\hline
  & \textsc{ChrF++} $\uparrow$  & \textsc{LLM-as-judge} $\uparrow$ & \textsc{LM Score} $\uparrow$ \\
\hline
\texttt{Mellum SFT} & $69.349 \pm 0.354$ & $0.676 \pm 0.003$ & $0.532 \pm 0.005$ \\
\texttt{Mellum Int. Base} & $75.256 \pm 0.349$ & $0.753 \pm 0.002$  & $0.643 \pm 0.005$ \\
\texttt{Mellum Int. DP} & $74.781 \pm 0.342$ & $0.730 \pm 0.003$ & $0.613 \pm 0.005$ \\
\hline
\end{tabular}
\label{tab:results}
\end{table}

\begin{figure}[t!]
    \centering
    \caption{Comparison of DP and non-DP models on training dynamics and privacy. \textbf{Left:} Validation loss and privacy budget ($\epsilon$) vs. data points seen. \textbf{Right:} ROC curve for MIA.}
    \begin{subfigure}{0.53\linewidth}
        \centering
        \includegraphics[width=\linewidth]{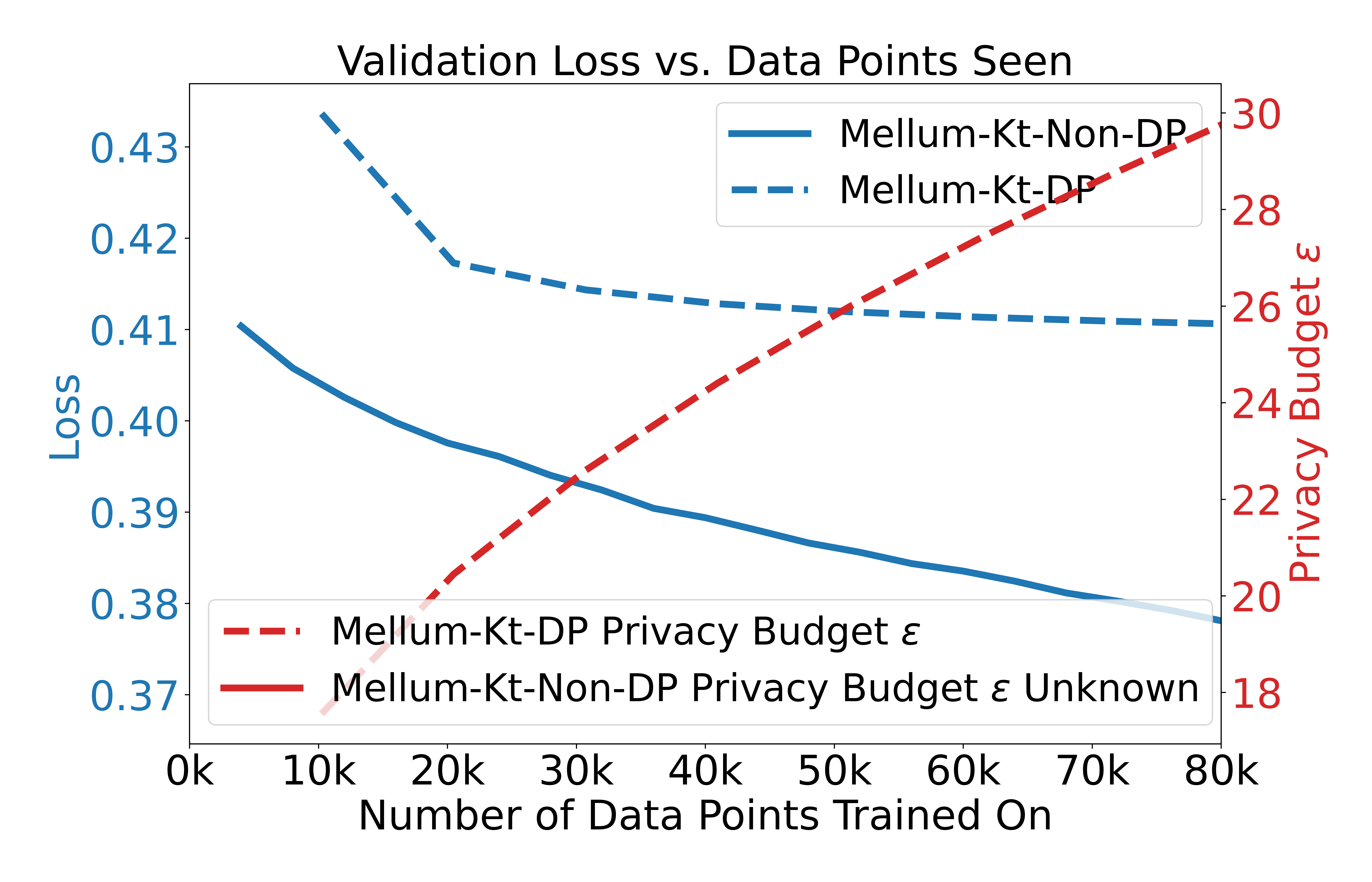}
        \label{fig:loss_vs_data}
    \end{subfigure}
    \begin{subfigure}{0.46\linewidth}
        \centering
        \includegraphics[width=\linewidth]{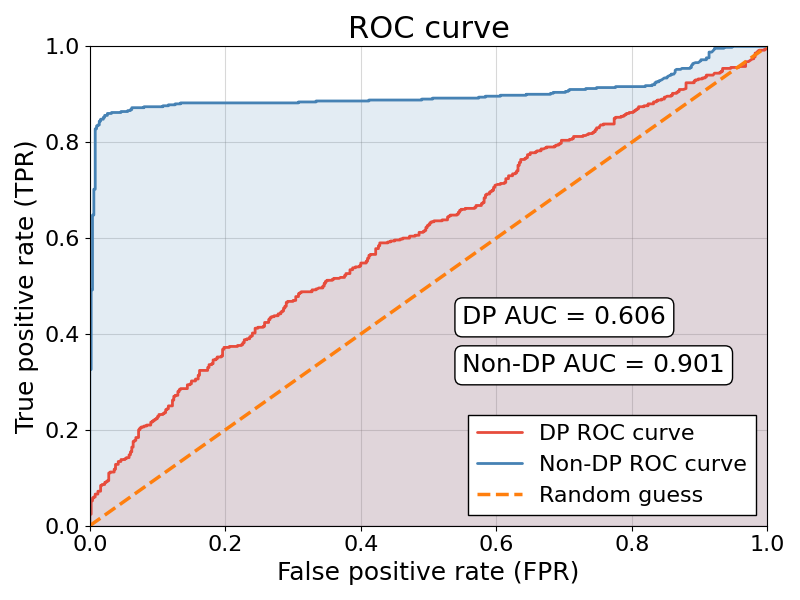}
        \label{fig:roc_mia}
    \end{subfigure}
    \label{fig:train_dynamics}
\end{figure}

In order to evaluate the model trained with DP, we conduct a two-fold analysis to measure its effectiveness in both privacy preservation and code generation utility, comparing it directly against the non-private baseline model. Our findings indicate that DP provides a powerful defense against privacy attacks while incurring a minimal and acceptable cost to model performance.

The primary goal of our work is to mitigate the privacy risks associated with training with user data. Figure~\ref{fig:train_dynamics} (right) clearly demonstrates the effectiveness of DP in achieving this objective. The non-private baseline model (\texttt{Mellum Int. Base}) is highly vulnerable to MIA, achieving an $AUC=0.901$. An AUC score this high indicates that an adversary can reliably distinguish between code snippets that were part of the model's training set and those that were not, posing a significant privacy risk.

In contrast, the differentially private model (\texttt{Mellum Int. DP}) effectively neutralizes this threat. Its ROC curve is nearly identical to the random guess baseline, resulting in an $AUC=0.606$. This score signifies that the attacker's ability to infer membership is close to flipping a coin, confirming that our DP training regimen successfully obfuscates the influence of individual training examples and provides a strong, practical defense against this class of privacy attacks.

\subsubsection*{Utility is Preserved Despite Strong Privacy Guarantees}
A critical consideration for deploying DP in practice is the trade-off between privacy and model utility. Our results, summarized in Table \ref{tab:results}, show that this trade-off is highly favorable in our use case.
The \texttt{Mellum Int. DP} model shows only a marginal decrease in the \textsc{ChrF++} score compared to the non-private baseline ($74.78$ vs. $75.26$) as well as for the other two metrics, a difference that is statistically insignificant. The DP model outperforms the \texttt{Mellum SFT} in all metrics. This high level of utility is particularly remarkable, given that the DP model was fine-tuned on only 80K internal data instances. This suggests that DP-SGD acts as a strong regularizer, potentially leading to better generalization on certain tasks even with significantly less data.

\subsubsection*{Analysis of Training Dynamics}
Figure \ref{fig:train_dynamics} (left) illustrates the training dynamics of both models. The validation loss for the DP model is consistently higher than that of the non-DP model. This is an expected and inherent characteristic of DP-SGD; the noise added to the gradients at each step to ensure privacy also slightly impedes the optimization process, resulting in a higher loss.

The same figure also plots the accumulation of the privacy budget ($\epsilon$) for the DP model over the course of training. As the model sees more data points, the privacy budget is gradually spent, increasing as more noisy gradient updates are performed. This visualization highlights the direct relationship between the amount of training and the privacy guarantee: more training requires a larger privacy budget. Our final model achieves has theoretical privacy guarantees with a final $\epsilon \approx 30$. While our final privacy budget exceeds the thresholds typically targeted in theoretical work ($\epsilon \le 10$), we demonstrate that it yields robust, practical protection within a real-world IDE environment. This finding aligns with prior research indicating that even looser privacy budgets effectively thwart MIA attempts to discern whether specific examples were included in the training data \cite{balle2022reconstructing, kaissis2023bounding, ziller2024reconciling}.

\section{Conclusion and Future Work}
\label{sec:conclusion}

This paper addresses the critical privacy risks, such as memorization and membership inference, that arise from fine-tuning LLMs on private user code for IDE autocomplete features. We present, to our knowledge, the first code completion LLM for an IDE that is fine-tuned with formal DP guarantees.

Our empirical results demonstrate a highly favorable privacy-utility trade-off. By fine-tuning the \textsc{Mellum} model using DP-SGD and LoRA, we provide a strong, practical defense against MIA, reducing the attack's success from a highly vulnerable to almost a random guess. Crucially, this robust privacy guarantee is achieved with a minimal impact on the model's utility.

Future work will involve exploring this trade-off between privacy and utility more deeply, particularly by experimenting with stricter privacy budgets (lower values of $\epsilon$) and their impact on model quality as well as training on more data. 

\begin{acks}
We thank Ivan Dolgov and Ivan Bondyrev for their technical support around evaluation and creating the dataset. Uladzislau Sazanovich for his continued support in this research direction. 
\end{acks}

\bibliographystyle{ACM-Reference-Format}
\balance
\bibliography{sample-base}

\end{document}